\documentclass[a4paper]{article}

\usepackage{icrc2013}

\title{Search for TeV $\gamma$-ray emission from AE Aqr coincident with high optical and X-ray states with the MAGIC telescopes} 

\shorttitle{AE Aqr observations with MAGIC}

\authors{

R.~L\'opez-Coto$^{1*}$, 
O. ~Blanch Bigas$^{1}$, 
J.~Cortina$^{1}$, 
D.~Hadasch$^{2}$, 
L. Takalo$^{3}$, 
D. Torres$^{2}$, 
on behalf of the MAGIC collaboration and
M.~Bogosavljevic$^{4}$, 
Z.~Ioannou$^{5}$, 
C.W.~Mauche$^{6}$,  
E.V. Palaiologou$^{7}$, 
M.A. P\'erez-Torres$^{8}$, 
T. Tuominen$^{3}$ 

}

\afiliations{
$^{1}$ IFAE, Edifici Cn., Campus UAB, E-08193 Bellaterra, Spain \\
$^{2}$ Institut de Ci\`encies de l'Espai (IEEC-CSIC), E-08193 Bellaterra, Spain \\
$^{3}$ Tuorla Observatory, University of Turku, FI-21500 Piikki\"o, Finland \\
$^{4}$ Astronomical Observatory Belgrade, 11060, Belgrade, Serbia\\
$^{5}$ Physics Department, College of Science, Sultan Qaboos University, P.O. Box 36, PC-123, Muscat, Oman\\
$^{6}$ Lawrence Livermore National Laboratory, 7000 East Ave, Livermore, CA 95125, USA \\
$^{7}$ Physics Department, University of Crete, P.O. Box 2208, GR-71003, Heraklion, Greece\\
$^{8}$ Instituto de Astrof\'{\i}sica de Andaluc\'{\i}a (CSIC), E-18080 Granada, Spain\\

}

\email{*rlopez@ifae.es}

\abstract{We report on observations of the nova-like cataclysmic variable AE Aqr performed by MAGIC. The observations were part of a quasi-simultaneous multi-wavelength campaign carried out between 2012 May and June covering the optical, UV, X-ray and gamma-ray ranges. MAGIC conducted the campaign and observed the source during 12 hours. The other instruments involved were KVA, Skinakas, and Vidojevica in the optic and Swift in the X-ray. We also used optical data from the AAVSO. The goals were to: monitor the variability of the source at different wavelengths, perform gamma-ray studies coincident with the highest states of the source at the other wavelengths, and confirm or rule out previous claims of detection of very-high-energy emission from this object. We report on a search for steady TeV emission during the whole observation, for variable TeV emission coincident with the highest optical and X-ray states and periodic TeV emission at the 33.08 s rotation period (30.23 mHz rotation frequency) of the white dwarf and its first harmonic (60.46 mHz rotation frequency). These are the first observations under good weather conditions performed by the present generation of IACTs for this object.}


\begin{document}
\maketitle

\section{Introduction}
Cataclysmic Variable stars (CVs) are binary systems comprising a white dwarf (WD) and another companion (usually a red dwarf) that transfers matter to the white dwarf. They are classified depending on the type of variation they manifest as novae, nova-like variables, dwarf novae, and magnetic CVs. They have outbursts observed at different wavelength, from radio to X-rays. TeV $\gamma$-ray emission from AE Aqr has also been reported by two different groups \cite{Narrabri92,Meintjes94} using the imaging atmospheric Cherenkov technique, but these detections could not be reproduced by the following generations of instruments using the same technique \cite{WhippleUL}.

AE Aqr is a bright nova-like cataclysmic variable star consisting of a magnetic white dwarf and a K4-5 V secondary. The orbital period of the system is $T_{\rm o}$=9.88 hours and the spin period of the white dwarf is $T_{\rm s}$=33.08 s \cite{Patterson79}, the shortest known. The system is located at a distance of $\sim$100 pc. It was originally classified as a DQ Her star \cite{Patterson94}, but it shows features that are not explained by this model, such as its violent activity at different wavelengths; the single-peaked Balmer emission lines, which produce Doppler tomograms that are not consistent with those of an accretion disk, and its fast spin-down rate. Its spin down rate was firstly measured to be \.P=5.64$\times10^{-14}$ s s$^{-1}$ \cite{deJager94}, although more recent X-ray measurements show that the spin-down rate is even larger \cite{MaucheBreak}. It exhibits flares 50 \% of the time, going in the visible regime from mag=12.5 (during quiescence) to mag=10 (during flares). These flares are emitted randomly all the time. Radio flares have been observed from this source, on timescales similar to the optical flares. In \cite{RelativisticElectrons} it is shown that the radio flares can be explained by relativistic electrons, which provides evidences of accelerated particles, radiating synchrotron emission in magnetized clouds. Hard X-rays have also been detected coming from the target with a 33 s-modulation \cite{MaucheBreak}. Due to this fact and the fast rotating period of the primary, there are WDs with similar behaviour as pulsars and they have been proposed to be contributors of low energy cosmic rays  \cite{SuzakuTerada}. It also has to be mentioned that the source was observed in 2005 in the context of a multiwavelength campaign by MAGIC and HESS, but due to bad weather conditions the data are not of good quality \cite{ MaucheMWL}.

Moreover, after the discovery of $\gamma$-ray emission from the symbiotic nova V407 Cygni \cite{ScienceFermi}, and the later report on emission from two additional novae \cite{ArxivFermi}, CVs have been included amongst high energy emitters. Due to the rapid rotation of its white dwarf, AE Aqr is one of the best candidates amongst the population of CVs to be observed by ground-based Cherenkov telescopes.

\section{Multiwavelength campaign}
\begin{figure}[t]
  \centering
  \includegraphics[width=0.5\textwidth]{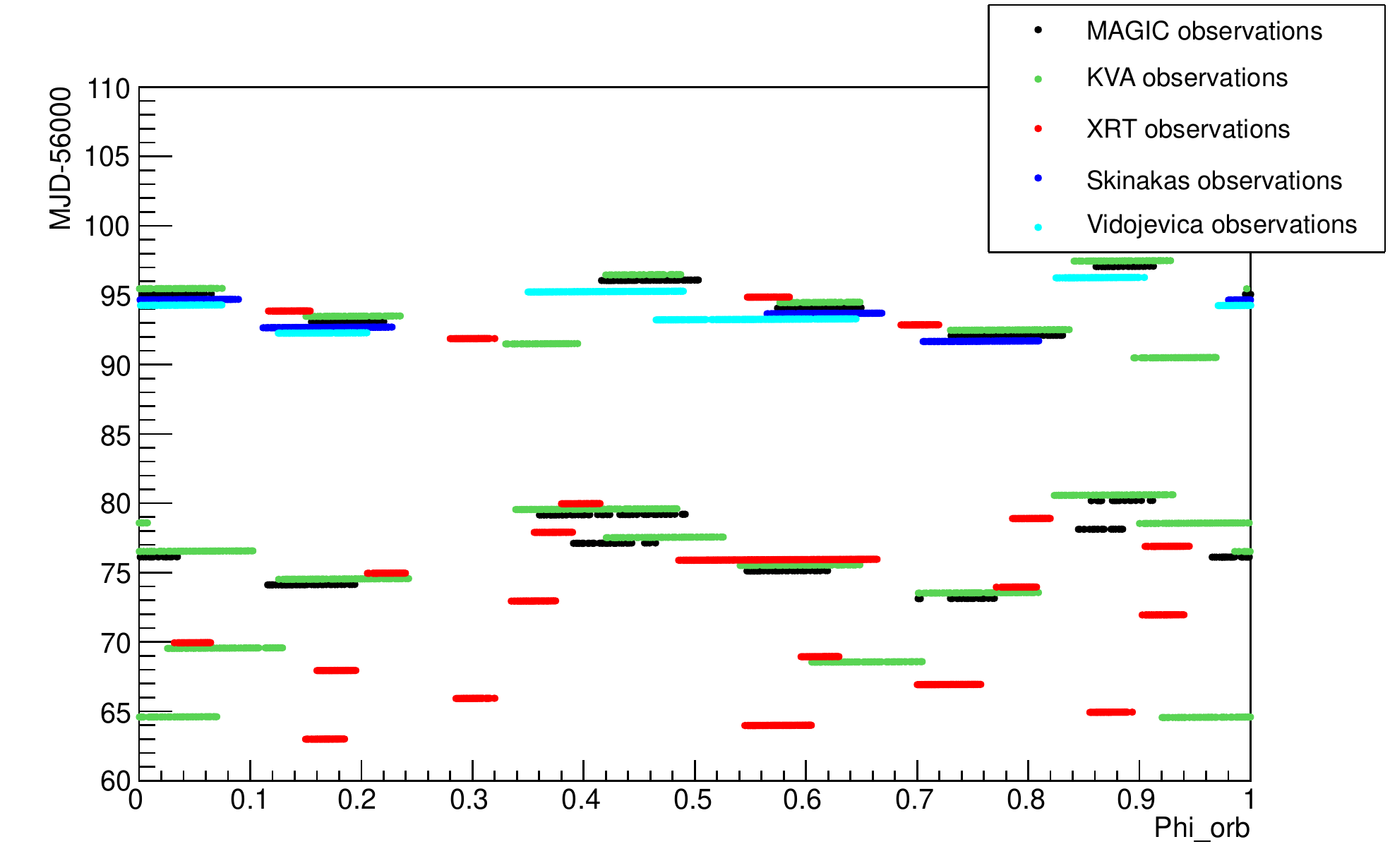}
  \caption{Summary of the multiwavelength campaign.  The instruments used, the Modified Julian Date and the orbital phases are plotted.}
  \label{campaign_summary}
 \end{figure}

During the period between 2012 May 15 and June 19, a multi-wavelength campaign triggered by the MAGIC telescopes was carried out to observe AE Aqr. The purpose of this campaign was to look for gamma-ray emission during the different states of the source at several wavelengths. In Figure \ref{campaign_summary} you can see a summary of the observations carried out during the campaign.

\subsection{MAGIC}
MAGIC is an Imaging Atmospheric Cherenkov telescope situated on the island of La Palma (28.8$^\circ$N, 17.9$^\circ$ W at 2225 m.a.s.l). It is a stereoscopic system that can achieve a sensitivity of (0.76 $\pm$ 0.03)\% of the Crab Nebula flux in 50 hours. The energy threshold achieved by the telescope for observations at low zenith angle is 50 GeV.

MAGIC observed AE Aqr during 15 non-consecutive nights during the period previously mentioned. The first night that weather conditions allowed observations, the telescope worked in stereo mode for 1.5 hours. After that, due to technical problems, the observations had to be performed in mono mode, accounting for a total of 10.5 hours. The source was observed at zenith angles ranging between 28$^\circ$ and 50$^\circ$.

\subsection{Swift}

Swift \cite{Swift} target-of-opportunity observations of AE Aqr were obtained during 25 orbits during 2012 May 15-June 2 and June 13-18. Data was obtained with the X-ray Telescope (XRT)\cite{XRT}, Ultraviolet/Optical Telescope (UVOT), and Burst Alert Telescope (BAT), although only the XRT data have been analyzed in support of the MAGIC observations. The XRT nominal exposures ranged from 37 s to 1553 s, with $\sim$950 s being typical. Excluding the shortest exposure and 3 exposures when the source was placed on one of the XRT detector dead stripes, there are 21 useful exposures for a total of 19.94 ks on-source. 

\subsection{Optical facilities}
Three optical telescopes participated in addition to the American Association of Variable Star Observers (AAVSO) in the campaign.

\subsubsection*{KVA telescopes}
The KVA optical telescopes are located on La Palma, but are
operated remotely from Finland. The two telescopes are attached
to the same fork. The larger telescope has a mirror diameter of
60 cm and the smaller 35 cm. The 35 cm telescope was used for simultaneous photometric
observations with MAGIC. The AE Aqr observations were performed in the B-band using 20 second exposures extending to $\sim$ 2 hours of data per night during 19 nights. The magnitude of the source was measured from CCD images using differential photometry. The comparison star magnitudes for AE Aqr from star 122 of the AAVSO AE Aqr finder chart were used.

\subsubsection*{Skinakas}
The data from the Skinakas Observatory in Crete were obtained with the 1.3-m Ritchey-Chr\'etien telescope located on the Skinakas mountain at an altitude of 1750 meters. The telescope has a focal ratio of f/7.6. The data were acquired with an Andor Tech DZ436 2048x2048 water cooled CCD. The physical pixel size is 13.5 microns resulting in 0.28 arcsec on the sky. The camera mode used was the 2 $\mu$s per pixel readout mode. At this mode the camera exhibits a readout noise of 8.14 electrons and its gain is 2.69 electrons/ADU. 

The observations were taken using a Bessel B filter using 10 second exposures, while the cycle time from the start of one exposure to the next was 14 seconds. Due to variable weather conditions, as well as the relatively low altitude of AE Aqr, the atmospheric seeing during the observations varied between 1-3 arcsec.

The data from Skinakas were taken during $\sim$ 1 hour for 4 nights.

The Skinakas data were reduced using IRAF routines. Differential photometry was performed using the photometry package DAOPHOT using 25 pixel circular apertures. For consistency with the data obtained by the  Vidojevica astronomical station 0.6-m telescope, as well as with previous photometric data obtained by the Skinakas telescope during the 2005 campaign (\cite{MaucheMWL}), the AE Aqr data were flux calibrated using the same comparison stars as before, namely the stars 122 and 124 in the AAVSO AE Aqr finder chart.

\subsubsection*{Vidojevica}
The Astronomical Station Vidojevica data have been obtained with the 60-cm Cassegrain telescope.
The telescope is used in the f/10 configuration with the Apogee Alta U42 CCD camera
(2048 x 2048 array, 13.5 micron pixels providing a 0.46 arcsec/pix plate scale).
The B filter from Optec Inc. (Stock No. 17446) was used for all observations.
The field centered on the target AE Aqr was observed continuously
 with 10 seconds of exposure time.
Only a fraction of the full CCD chip field of view, roughly 5 arcmin on a side,
was readout incurring for approximately 4 seconds of readout time, and
14 seconds total cycle time between exposures.

The data were taken for periods between $\sim$1-2 hours for 5 nights.

The data reduction was performed using standard procedures in IRAF.
The photometry was performed with Source Extractor,
using 10 pixel (4.6") diameter circular apertures.
Typical seeing conditions during the observations were 2" FWHM.
Target AE Aqr flux was calibrated using the same comparison stars as for Skinakas

\newpage

\subsubsection*{AAVSO}
A number of AAVSO observers provided us with additional observations. However, due to the low time resolution of these observations we did not include the AAVSO data into our timing an AAVSO data into our timing analysis

\

\section{Results}

\begin{figure}[]
 \centering
 \includegraphics[width=0.5\textwidth]{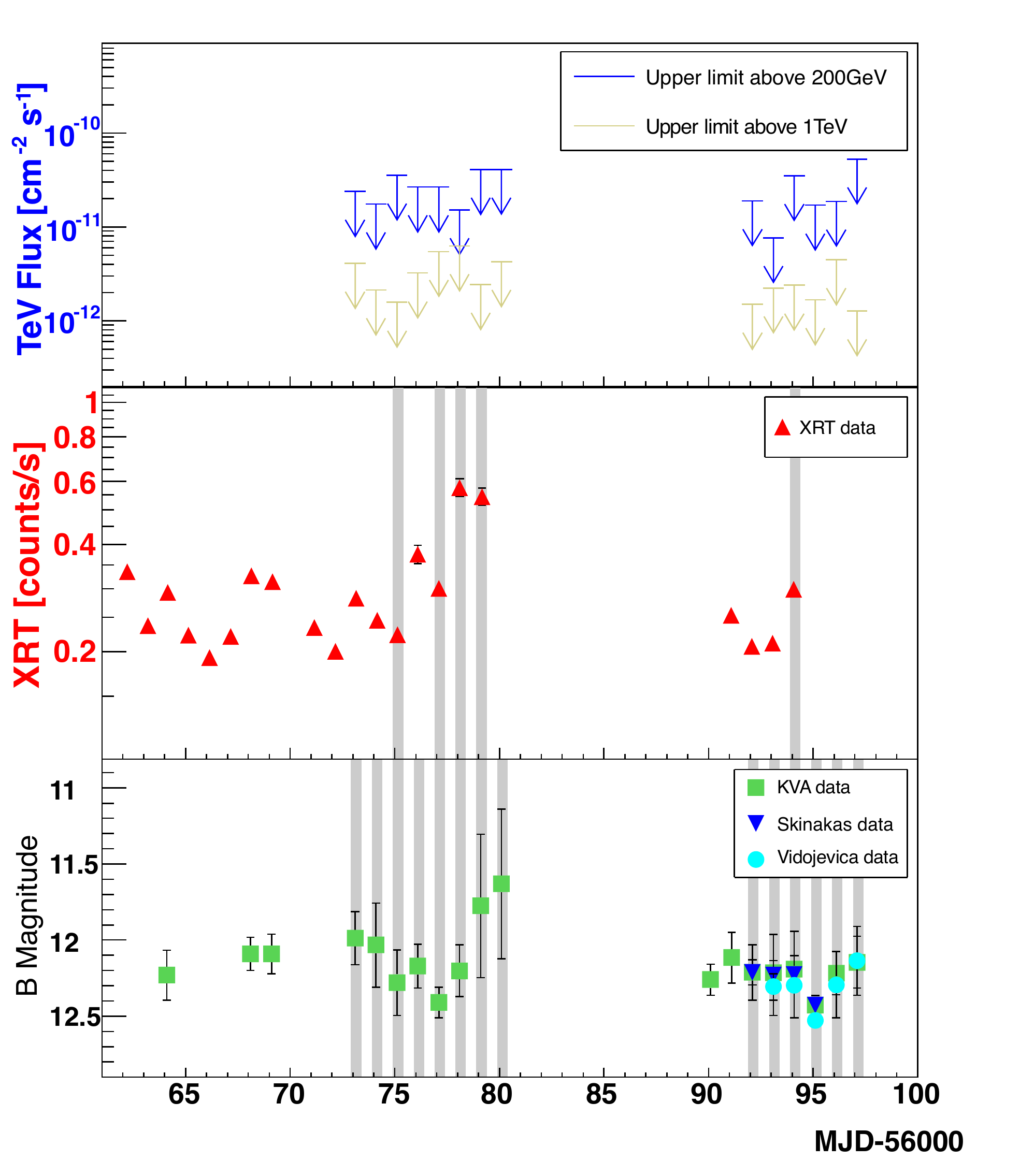}
\caption{Light curves of the multiwavelength campaign. The plot includes MAGIC daily upper limits considering a power-law spectrum with a -2.6 slope (top), XRT flux (medium) and B magnitude measured by the optical telescopes (bottom). For the optical data, as the source variability is very large, the error bars indicate the maximum and minimum state reached during that observation night. The shadowed areas indicate the observations simultaneous with MAGIC.
}
  \label{LC}
 \end{figure}

The results of the campaign are shown in this section. The light curves of the multi-wavelength campaign can be seen in Fig. \ref{LC}.

\

\subsection{MAGIC results}
MAGIC data have been analyzed in different ways in order to look for a steady signal or a periodic one. A summary of the MAGIC observations can be found in Table \ref{tab.2}. Most of those observations were simultaneous to the optical and X-ray ones. Therefore, a study of the correlation of optical/X-ray flux with the possible $\gamma$-ray emission has been done.

\begin{table}[tb]
\begin{center}
\begin{tabular}{|c|c|}

\hline
    Date   &	MAGIC observation \\
 & time window \\
\hline
\hline
   19/05/12 & 03:12 - 04:51 \\
\hline  
   26/05/12 & 02:47 - 03:33 \\
\hline
   27/05/12 & 02:40 - 03:25  \\
\hline
   28/05/12 & 02:40 - 03:23  \\
\hline
   29/05/12 & 02:34 - 03:17  \\
\hline
   30/05/12 &  02:28 - 03:14 \\
\hline
   31/05/12 &  02:15 - 03:09 \\
\hline
   01/06/12 & 03:37 - 04:57  \\
\hline
   02/06/12 &  03:42 - 04:54 \\
\hline
   14/06/12 & 01:32 - 02:34 \\
\hline
   15/06/12 &  01:31 - 02:10 \\
\hline
   16/06/12 & 01:23 - 02:06 \\
\hline
   17/06/12 & 01:18 - 01:58 \\
\hline
   18/06/12 & 01:13 - 02:04 \\
\hline
   19/06/12 & 01:22 - 01:55 \\
\hline

\end{tabular}
\caption{Observation time for every night (all the times are given in UT) for the MAGIC observations of AE Aqr performed during the 2012 multi-wavelength campaign.}
\label{tab.2}
\end{center}
\end{table}

\subsubsection*{Steady emission}

First of all, we stacked all the data together and look for a $\gamma$-ray significant signal. Since these observations took place at zenith angles $>$ 28$^\circ$, the energy threshold of the telescope was 250 GeV. As the shape of the spectral distribution is not known, we have considered power-laws with different spectral indexes, as well as different energy thresholds in the upper limits calculation. We did not find a significant signal in any energy bin. The upper limits for the steady emission of the source considering power-law spectra with different slopes can be found in Table \ref{Slopes}. We also computed daily integral upper limits for the steady emission of the source considering a power-law with a -2.6 slope (Crab-like) spectrum. Those upper limits are shown in Fig. \ref{LC}.

\begin{table}[!htb]
\begin{center}
\begin{tabular}{|c|c|c|}

\hline
  &  \multicolumn{2}{|c|}{MAGIC Integral upper limits}\\
  Slope  & \multicolumn{2}{|c|}{$[$cm$^{-2}$s$^{-1}]$}    \\
   \cline{2-3}
 & above 250 GeV & above 1TeV \\
\hline    
 -2   & 4.165e-12& 7.574e-13 \\
 -2.6    & 6.391e-12  & 7.401e-13\\
  -3   & 8.011e-12 & 7.377e-13  \\
\hline
\end{tabular}
\end{center}
\caption{MAGIC upper limits considering a power-law spectrum with different slopes}\label{Slopes}
\end {table}

\subsubsection*{Emission coincident with several optical states}
According to the purpose of the multi-wavelength campaign, we have studied the behavior of the source coincident with different optical states. We selected the events coincident with B magnitude $<$ 12 and B magnitude $<$ 11.5. The integral upper limits for those states can be found in Table \ref{UL_optical}.

\begin{table}[!htb]
\begin{center}
\begin{tabular}{|c|c|c|}

\hline
  B Magnitude &  \multicolumn{2}{|c|}{MAGIC Upper limits}\\
   & \multicolumn{2}{|c|}{$[$cm$^{-2}$s$^{-1}]$}    \\
 \hline 
 & above 250 GeV & above 1TeV\\
\hline    
Brighter than 11.5   & 2.051e-11 & 1.624e-12 \\
Brighter than 12    & 7.299e-12 & 1.189e-12\\
\hline
\end{tabular}
\end{center}
\caption{MAGIC upper limits for different optical states}\label{UL_optical}
\end {table}

\subsubsection*{Pulsed emission}
A search for pulsed emission at the spin frequency of the white dwarf and its first harmonic has also been performed. We used the ephemeris for the source presented in  \cite{MaucheBreak} and limit the signal region to the bin corresponding to the maximum of the XRT spin-phase-folded light curve (see Fig. \ref{XRT_spin}). We did not find any hint of periodic signal for either frequencies.

  \begin{figure}[ht]
  \centering
  \includegraphics[width=0.5\textwidth]{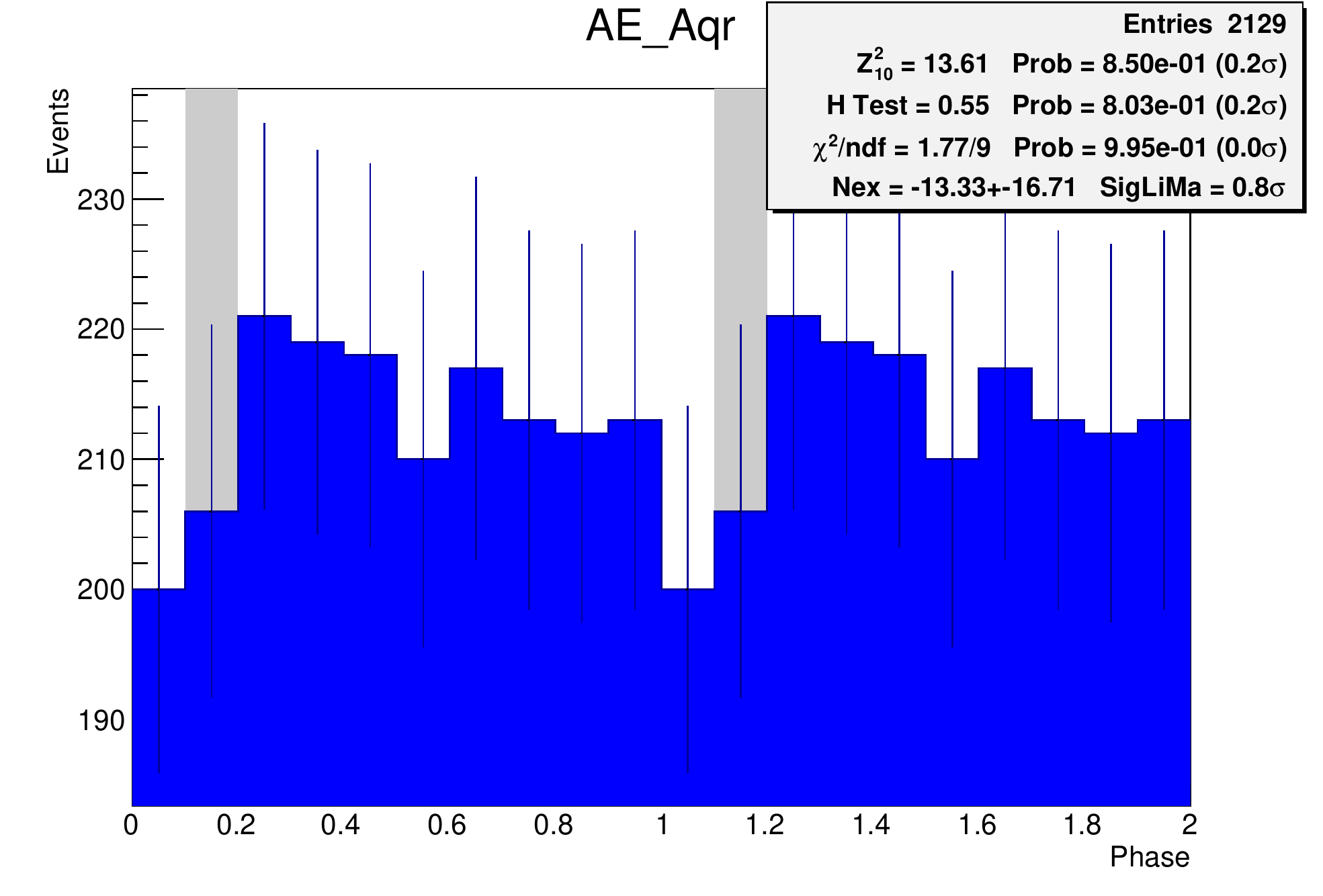}
  \caption{Phaseogram for the MAGIC data above 250 GeV for a frequency of 30.23mHz}
  \label{Phaseogram}
 \end{figure}

\subsection{Swift results}

\begin{figure}[ht]
  \centering
  \includegraphics[width=0.5\textwidth]{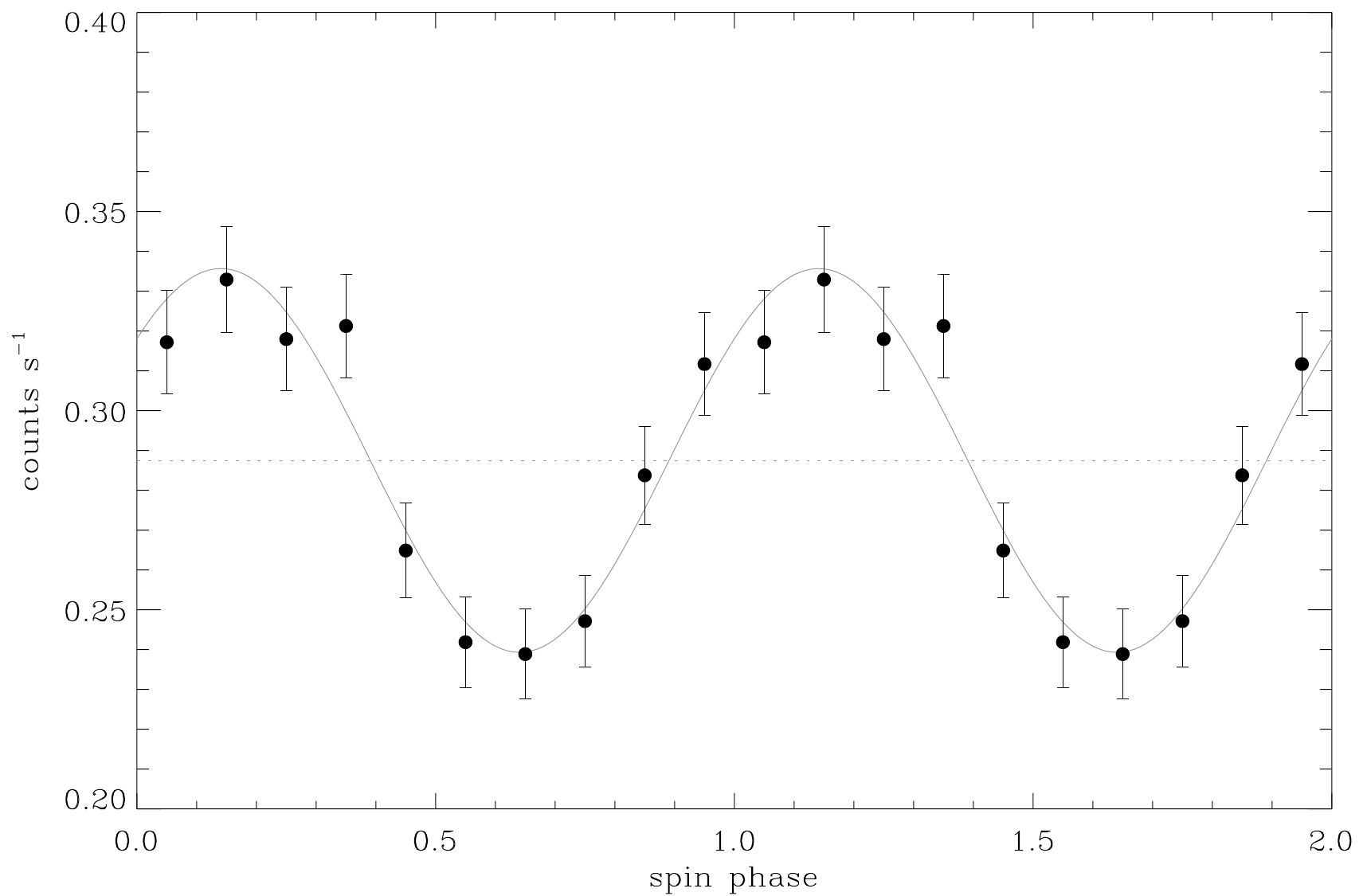}
  \caption{XRT spin-phase-folded light curve}
  \label{XRT_spin}
 \end{figure}

The XRT event data was used to produce the X-ray light curve shown in Fig. \ref{LC} and the spin-phase-folded light curve shown in Fig. \ref{XRT_spin}. 
 
 The spin-phase-folded light curve is fit with a cosine function A($\phi_{spin}$)=A$_0$+A$_1$cos[2$\pi$($\phi_{spin}$-$\phi_{off}$)] with constants:

\begin{eqnarray*}
 \mbox{A}_{0} = 0.287 \pm 0.003\\
 \mbox{A}_{1} = 0.048 \pm 0.005\\
 \phi_{off} = 0.141 \pm 0.018
\end{eqnarray*}

with $\chi^2$/dof=4.94/7=0.71. Hence, the relative pulse amplitude A$_1$/A$_0$=16.8\%$\pm$1.9\%, which is comparable to, but higher than, that measured by ASCA, XMM-Newton, and Chandra, which are 12.9\%, 9.5\%, and 15\%, respectively (see Table 2 of \cite{MaucheBreak}).

Because of $\phi_{off}$ is not equal to zero, the \cite{MaucheBreak} cubic ephemeris is not exact and these and other data should be used to update and refine it.

\subsection{Optical results}
In the optical part we have plotted the results of all the observations together in order to check the consistency between the magnitudes measured by the different telescopes. The simultaneity of the observations let us determine the state of the source at TeV energies at different optical emission levels. The highest optical state was achieved the night of the 2nd of June, reaching B$_{mag}$=11.08. The spin-phase-folded light curve could not be produced due to the low time resolution of the observations.

\section{Conclusion}

We have carried out the most sensitive observations by an IACT of a cataclysmic variable star. The very-high-energy observations go together with optical and X-ray observations that help to characterize the behavior of the source at different states. With the X-ray data we will be able to update the ephemeris of the source. We have searched for steady $\gamma$-ray emission during the whole observation period, coincident with different optical states and pulsed  $\gamma$-ray emission as well. We do not find any significant emission from AE Aqr in any of the searches performed.

\vspace*{0.5cm}
\footnotesize{{\bf Acknowledgment:}{ We would like to thank the Instituto de Astrof\'{\i}sica de
Canarias for the excellent working conditions at the
Observatorio del Roque de los Muchachos in La Palma.
The support of the German BMBF and MPG, the Italian INFN, 
the Swiss National Fund SNF, and the Spanish MINECO is 
gratefully acknowledged. This work was also supported by the CPAN CSD2007-00042 and MultiDark
CSD2009-00064 projects of the Spanish Consolider-Ingenio 2010
programme, by grant 127740 of the Academy of Finland, by the DFG Cluster of Excellence ``Origin and Structure of the 
Universe'', by the DFG Collaborative Research Centers SFB823/C4 and SFB876/C3,
and by the Polish MNiSzW grant 745/N-HESS-MAGIC/2010/0. 
CWM's contribution to this work was performed under the auspices of the U.S. Department of Energy by Lawrence Livermore National Laboratory under Contract DE-AC52-07NA27344. MB acknowledges support of Serbian MESTD through grant ON176021. The authors thank N. Gehrels for approving our request for target-of-opportunity observations and the Swift Science Operations Team for scheduling them.
We would also like to thank the American Association of Variable Star Observers for their help during the campaign.}}


\end{document}